\documentclass[aip, ajp, reprint]{revtex4-2} % Two column
\usepackage{amsmath}  % needed for \tfrac, \bmatrix, etc.
\usepackage{amsfonts} % needed for bold Greek, Fraktur, and blackboard bold
\usepackage{graphicx} % needed for figures
\usepackage{siunitx} %\SI[separate-uncertainty = true]{1.0(1)e-3}{\second}
\usepackage{verbatim}
\usepackage{color}
\usepackage{mathrsfs}
\usepackage{multirow}

\begin{document}

% Be sure to use the \title, \author, \affiliation, and \abstract macros
% to format your title page.  Don't use lower-level macros to  manually
% adjust the fonts and centering.

\title{Transients in lossy transmission lines}
% In a long title you can use \\ to force a line break at a certain location.

%\author{S. D. Delichte}
%\author{Y. J. Lu}

\author{J. S. Bobowski}
\email{jake.bobowski@ubc.ca} % optional
%\altaffiliation[permanent address: ]{101 Main Street, 
%  Anytown, USA} % optional second address
% If there were a second author at the same address, we would put another 
% \author{} statement here.  Don't combine multiple authors in a single
% \author statement.
\affiliation{Department of Physics, University of British Columbia, Kelowna, British Columbia, Canada V1V 1V7}
% Please provide a full mailing address here.

% See the REVTeX documentation for more examples of author and affiliation lists.

\date{\today}

\begin{abstract}
Numerical inverse Laplace transforms are used to analyze the transient response of non-ideal transmission lines to a voltage step.  We find that the detailed shape of the transient response is sensitive conductor losses, but insensitive to dielectric losses.  Furthermore, we find that the low-loss approximations for the transmission line propagation constant and characteristic impedance in the complex-frequency domain are sufficient to accurately model the observed transient response.  We also investigate the effects of: (1) a parasitic capacitance terminating the open end of the coaxial transmission line, (2) the input impedance of the oscilloscope used to make the measurements, and (3) the finite rise time of the voltage step.  Finally, we cool a semi-rigid coaxial cable in liquid nitrogen so as to reduced conductor lossless and observe a transient response that is closer to that expected from an ideal lossless line.
\end{abstract}
% AJP requires an abstract for all regular article submissions.
% Abstracts are optional for submissions to the "Notes and Discussions" section.

\maketitle % title page is now complete

\section{INTRODUCTION}
In a recent paper we investigated the non-ideal characteristics of transmission lines using frequency-domain insertion loss measurements.\cite{Bobowski:2020}  Analyses of the data yielded the frequency dependencies of the line's per-unit-length resistance due to conductor losses and per-unit-length conductance due to dielectric losses.  As a precursor to the frequency-domain measurements, an analysis of the line's transient response to a voltage step was carried out to determine its per-unit-length capacitance and inductance.  That analysis assumed an ideal (lossless) transmission line and was unable to account the for non-ideal features that appeared in the staircase-like time-domain data.  These features include rounding of the step corners and non-zero slopes of the `horizontal' sections of the steps.\cite{Bobowski:2020}  

In this paper, we use the Laplace transform formalism to analyze the transient response of non-ideal transmission lines to a voltage step.  The approach is to first analyze the problem in the complex-frequency, or $s$-domain, using standard circuit analysis techniques.\cite{Haus:1989, Pozar:2012, Collier:2013, Snoke:2015}  We then evaluate the inverse Laplace transform to deduce the desired transient response.\cite{Shenkman:2005}  The required inverse Laplace transforms do not have analytical solutions and must be evaluated numerically.\cite{Griffith:1990, Valsa:1998}  We apply numerical methods developed by Valsa and Bran\v{c}ik which can by applied to rational, irrational, and transcendental functions of $s$ with high accuracy.  In particular, the method is well-suited to ``lossy transmission lines with frequency-dependent parameters'' which is precisely the case that we wish to consider.\cite{Valsa:1998}  

The numerical routine described in Ref.~\onlinecite{Valsa:1998} was implemented as a MATLAB function called INVLAP written by Valsa and posted to the MathWorks File Exchange.\cite{Valsa:2011}  Our inverse Laplace transforms were evaluated by calling the INVLAP function.  The accuracy of numerical calculation is determined by the parameters $a$, $n_\mathrm{sum}$, and $n_\mathrm{dif}$ which the user supplies as arguments when calling the function.  As described in Ref.~\onlinecite{Valsa:1998}, the $a$ parameter is used when writing an approximation for the factor $e^{st}$ that appears in the inverse Laplace transform of the function $F(s)$
\begin{equation}
f(t)=\mathscr{L}^{-1}\{F(s)\}=\frac{1}{2\pi j}\int\limits_{\alpha-j\infty}^{\alpha+j\infty} F(s)e^{st}ds,
\end{equation}
where $j=\sqrt{-1}$.  Following the recommendation of the authors of the method, we used $a=6$ in our implementations of INVLAP.\cite{Valsa:2011}  The parameter $n_\mathrm{sum}$ determines the number of terms to keep in the infinite sum used to approximate $e^{st}$.  We used $n_\mathrm{sum}=8000$.  Finally, $n_\mathrm{dif}$ determines the number of terms to keep in a second sum that is used to estimate the error caused by truncating the first sum.  This second sum converges quickly and we used $n_\mathrm{dif}=160$.\cite{Valsa:1998} 

\begin{figure*}
\centering{
(a)~\includegraphics[keepaspectratio, width=1.1\columnwidth]{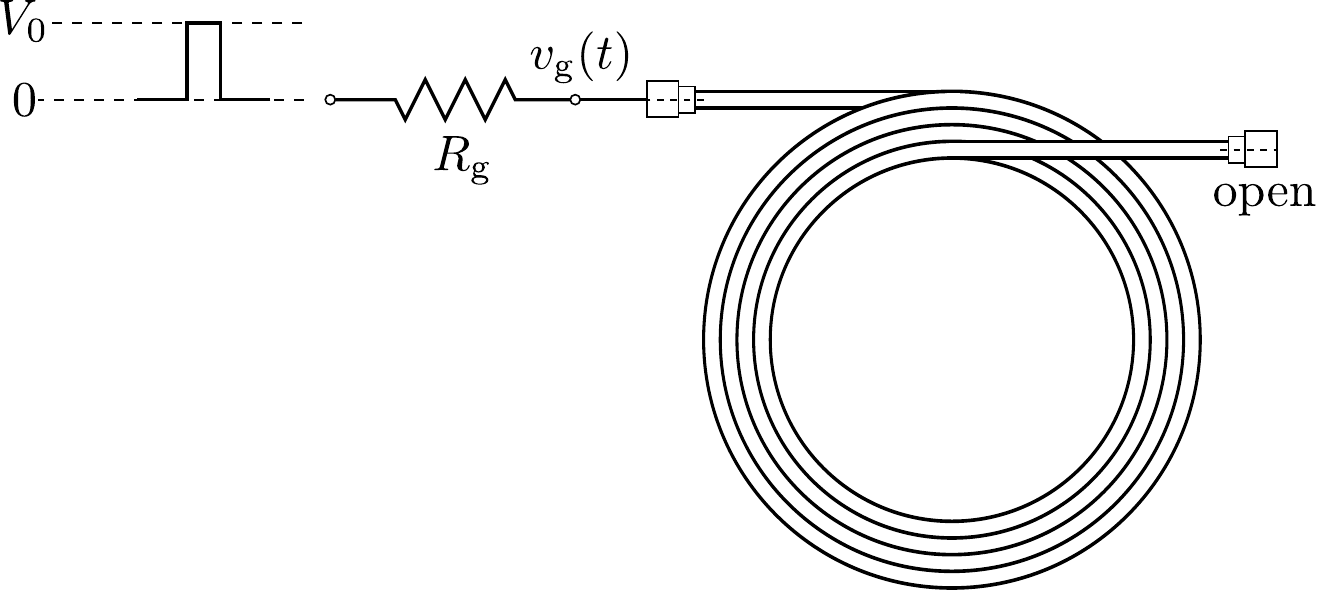}\qquad
(b)~\raisebox{0.08\height}{\includegraphics[keepaspectratio, width=0.6\columnwidth]{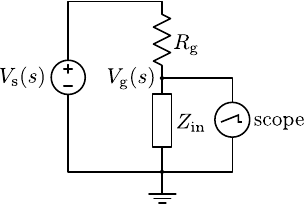}}
}
\caption{\label{fig:transients}(a) Schematic diagram of the experimental setup used to measure the transient response of the transmission line.
(b) The $s$-domain equivalent circuit of (a).  $V_\mathrm{s}$ is the Laplace transform of the voltage pulse and $V_\mathrm{g}$, measured using an oscilloscope, is the voltage at the node between $R_\mathrm{g}$ and the transmission line.   
}
\end{figure*}

\section{COMPLEX-FREQUENCY DOMAIN ANALYSIS}
The input impedance of a lossy transmission line of length $\ell$ terminated by a load impedance $Z_\mathrm{L}$ is given by
\begin{equation}
Z_\mathrm{in}=Z_\mathrm{c}\frac{Z_\mathrm{L} + Z_\mathrm{c}\tanh\left(\gamma\ell\right)}{Z_\mathrm{c} + Z_\mathrm{L}\tanh\left(\gamma\ell\right)},\label{eq:Zin}
\end{equation}
where $Z_\mathrm{c}$ and $\gamma$ are the characteristic impedance of the line and the signal propagation constant, respectively.  In terms of the fundamental equivalent circuit parameters, the characteristic impedance and propagation constant are given by
\begin{align}
Z_\mathrm{c}&=\sqrt{\frac{R + sL}{G + sC}},\label{eq:Zc}\\
\gamma&=\sqrt{\left(R + sL\right)\left(G + sC\right)}.\label{eq:gamma}
\end{align}  
In these expressions, $L$ and $C$ are the per-unit-length capacitance and inductance of the transmission line, respectively.  $R$ is the per-unit-length resistance due to conductor losses and $G$ is the per-unit-length conductance due to dielectric losses.  The variable $s$ is the complex frequency which is typically given by $j\omega$, where $\omega$ is angular frequency.\cite{Haus:1989, Pozar:2012, Collier:2013, Snoke:2015, Shenkman:2005}

In the low-loss limit, for which $R/\left(sL\right) + G/\left(sC\right)\ll 1$, the characteristic impedance and propagation constant can be approximated as
\begin{align}
Z_\mathrm{c} &\approx Z_0\left(1 + \frac{v_0}{2s}k_-\right),\label{eq:ZcApprox}\\
\gamma &\approx \frac{s}{v_0}\left(1 + \frac{v_0}{2s}k_+\right),\label{eq:gammaApprox}
\end{align}
where $Z_0=\sqrt{L/C}$, $v_0=1/\sqrt{LC}$, and $k_\pm= R/Z_0\pm GZ_0$.\cite{Bobowski:2020}

The experimental setup used to study the transmission line transient response is shown schematically in Fig.~\ref{fig:transients}(a).  A voltage pulse is applied to one end of a resistor $R_\mathrm{g}$ whose opposite end is connected to the center conductor of a long coaxial cable.  The duration of the pulse is long compare to the time that it takes a signal to travel the length of the cable.  The other end of the coaxial cable is left open.  The voltage $v_\mathrm{g}$ at the junction between $R_\mathrm{g}$ and the line is measured as a function of time using an oscilloscope.  The equivalent $s$-domain circuit of the experimental setup is shown in Fig.~\ref{fig:transients}(b).  $V_\mathrm{s}(s) = \mathscr{L}\{v_\mathrm{s}(t)\}$ and $V_\mathrm{g}(s) = \mathscr{L}\{v_\mathrm{g}(t)\}$ are the Laplace transforms of $v_\mathrm{s}$ and $v_\mathrm{g}$, respectively, and $\mathscr{L}\{f(t)\}=\int_0^\infty f(t)e^{-st}dt$.  $Z_\mathrm{in}$ represents the input impedance of the transmission line and is given by Eq.~(\ref{eq:Zin}).  The circuit shown in Fig.~\ref{fig:transients}(b) is a voltage divider and, if it is assumed that the oscilloscope has a very high input impedance, $V_\mathrm{g}$ is given by
\begin{equation}
V_\mathrm{g}=V_\mathrm{s}\frac{Z_\mathrm{in}}{R_\mathrm{g} + Z_\mathrm{in}}.\label{eq:Vg}
\end{equation} 

\begin{figure}
\centering{
\includegraphics[width=0.9\columnwidth]{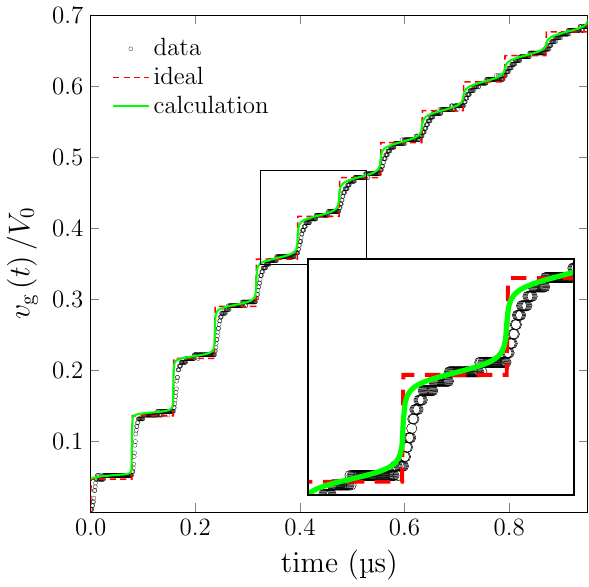}
}
\caption{\label{fig:Ronly}Transient response of a \SI{8.07}{\meter} length of UT-141 coaxial cable when using $R_\mathrm{g}=\SI{1}{\kilo\ohm}$.  The plot includes the experimentally-measured response (data points), the calculated response of a lossless transmission line (dashed line), and the calculated response of a low-loss line (solid line).
}
\end{figure}

\section{TRANSIENT RESPONSE}\label{sec:transient}
This section presents the results of the numerical inverse Laplace transforms used to determine the time dependence of $v_\mathrm{g}$.  Various forms of $Z_\mathrm{in}$ and $V_\mathrm{s}$ are considered and numerical results are compared to experimental data.

\subsection{Low-loss approximation with negligible conductance}\label{sec:Ronly}
The first case considered is the low-loss limit given by Eqs.~(\ref{eq:ZcApprox}) and (\ref{eq:gammaApprox}) with $k_+ = k_- \approx R/Z_0$.  Furthermore, since $Z_\mathrm{L}$ is an open circuit, $Z_\mathrm{in}\approx Z_\mathrm{c}\coth\left(\gamma\ell\right)$.  Finally, if we take $v_\mathrm{s}$ to be an instantaneous voltage step from zero to $V_0$ at $t=0$, then $V_\mathrm{s}=V_0/s$.  Combining all of these results yields
\begin{equation}
V_\mathrm{g} = \frac{V_0}{s}\left[\frac{1}{\left(R_\mathrm{g}/Z_\mathrm{c}\right)\tanh\left(\gamma\ell\right)+1}\right],\label{eq:lowloss}
\end{equation}
where $Z_\mathrm{c}\approx Z_0\left[1+\left(\tau s\right)^{-1}\right]$, $\gamma\approx\left(s/v_0\right)\left[1+\left(\tau s\right)^{-1}\right]$, and \mbox{$\tau\equiv 2Z_0/\left(v_0R\right)$}.

Figure~\ref{fig:Ronly} shows the transient response of a \SI{8.07}{\meter} length of UT-141 semi-rigid coaxial cable.  Also shown are the responses calculated using inverse Laplace transforms.  Equation~(\ref{eq:lowloss}) can be inverted analytically for the case of an ideal lossless line with $R=0$ ($\tau\to\infty$).  The result of that calculation using $Z_0=\SI{49.05}{\ohm}$ and $v_0/c=0.6795$, where $c$ is the vacuum speed of light, is shown as the dashed line in Fig.~\ref{fig:Ronly}.  These parameter values do a reasonable job of capturing the general shape of the experimental data and are close to the values extracted from the analysis presented in Ref.~\onlinecite{Bobowski:2020}.  However, the ideal response does not reproduce many of the detailed features of the experimental data which include rounded step corners, finite rise times for each step, and non-zero slopes between steps.  All of these non-ideal features become more exaggerated at time evolves.  We note also that we used INVLAP to calculate the response of a lossless line numerically and it produced a solution that was nearly indistinguishable from the analytical solution. 

To include non-zero conductor losses, we used the $s$-dependence of $R$ implied by the frequency-domain measurements reported in Ref.~\onlinecite{Bobowski:2020}.  In that work, we found that the effective conductor resistance could be modeled as $R=a\sqrt{f}$, where the square root frequency dependence is due to the electromagnetic skin depth of the conductors.  For a UT-141 coaxial cable with a copper outer conductor and a silver-plated copperweld (SPCW) center conductor, the value of the coefficient was experimentally determined to be $a=\SI{1.252e-4}{\ohm\,\second\tothe{1/2}/\meter}$.  The $s$-dependence of $R$ is determined by writing $f=-js/\left(2\pi\right)$ which leads to
\begin{equation}
R=ae^{-j\pi/4}\sqrt\frac{s}{2\pi}=\frac{a}{2}\left(1-j\right)\sqrt{\frac{s}{\pi}}.
\end{equation}
Using this model for $R$, the inverse Laplace transform of Eq.~(\ref{eq:lowloss}) was evaluated numerically and the results, using the same $Z_0$ and $v_0$ parameters as above, are shown using the solid line in Fig.~\ref{fig:Ronly}.  The calculated response has the desired rounded step corners and non-zero slopes between steps, however, the transition time from one step to the next is still much smaller than observed in the experimental data.  This difference arises because the calculated transient response assumed an instantaneous voltage step $v_\mathrm{s}$.  In the experimental system, the source voltage transitions from zero to $V_0$ over a finite time interval.  We consider voltage steps with finite rise times in Sec.~\ref{sec:finiteRise}.

We also calculated the transient response when including a non-zero conductance due to dielectric losses.  The $s$-dependence of $G$ from frequency-domain measurements was found to be $G= -jbs/2\pi$ with \mbox{$b=\SI{1.31e-13}{\second\ohm\tothe{-1}\meter\tothe{-1}}$} for the UT-141 semi-rigid coaxial cable.\cite{Bobowski:2020}  We found that including dielectric losses resulted is almost no change to the calculated response shown in Fig.~\ref{fig:Ronly} which included only conductor losses.

\begin{figure*}
\centering{
(a)~\includegraphics[width=0.9\columnwidth]{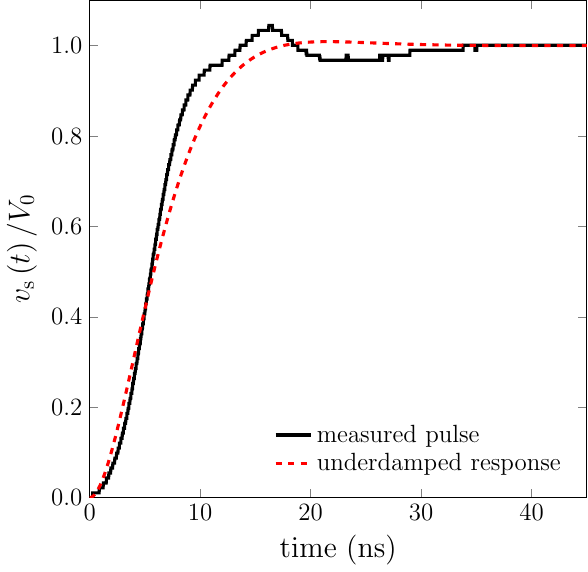} \qquad (b)~\includegraphics[width=0.9\columnwidth]{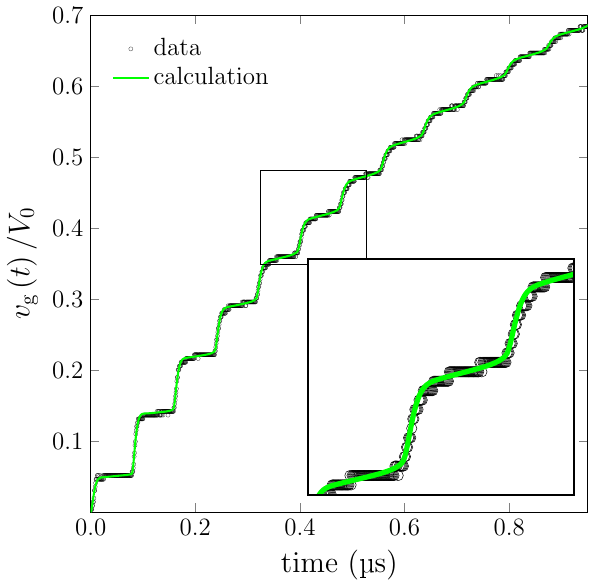} 
}
\caption{\label{fig:finiteRise}(a) The solid line shows the rise time of the pulse output by the HP 8011A pulse generator when connected to $R_\mathrm{g}=\SI{1}{\kilo\ohm}$ in series with a \SI{8.07}{\meter} length of UT-141 semi-rigid coaxial transmission line. The dashed line shows the transition of an underdamped oscillator ($f_0=\SI{42}{\mega\hertz}$, $Q=0.6$) between two equilibrium voltages (zero and $V_0$). (b) The measured and calculated transient response of a \SI{8.07}{\meter} long coaxial cable.  The calculated response includes conductor losses and the finite rise time of the input voltage pulse.}
\end{figure*}

\subsection{Voltage steps with finite rise times}\label{sec:finiteRise}
All practical pulse generates transition from zero to a nonzero voltage over a finite time.  The solid line in Fig.~\ref{fig:finiteRise}(a) shows the rise time of the voltage pulse measured at the node to the left of $R_\mathrm{g}$ in Fig.~\ref{fig:transients}(a).  Our pulse was supplied by an HP 8011A pulse generator and the measurement was made using a Tektronix TBS 1104 digital oscilloscope.  This measurement reveals a non-negligible rise time of approximately \SI{10}{\nano\second} followed by a damped oscillation before settling at an equilibrium voltage of $V_0$.    

To take into account this effect in our transient response calculations, we generalized our model for $V_\mathrm{s}$.  The inverse Laplace transform of $V_\mathrm{s}=V_0/s$ represents a voltage step that transitions from zero to $V_0$ instantaneously at $t=0$.  One can instead model the pulse generator as a damped oscillator with a resonant frequency $f_0$ and quality factor $Q$.  For an underdamped oscillator ($Q>0.5$), the time to transition between equilibrium states is approximately given by $2Q/\omega_0$, where $\omega_0=2\pi f_0$.  $V_s$ for an underdamped oscillator in the $s$-domain is given by 
\begin{equation}
\frac{V_\mathrm{s}}{V_0}=\frac{\omega_0^2}{s\left(s^2+\left(\omega_0/Q\right)s+\omega_0^2\right)}.\label{eq:VsV0}
\end{equation}
Evaluating the inverse Laplace transform Eq.~(\ref{eq:VsV0}) results in the time-domain response
\begin{multline}
v_s(t)/V_0=\mathscr{L}^{-1}\{V_\mathrm{s}(s)/V_0\}\\
 =1-e^{-\omega_0t/\left(2Q\right)}\left(\cos\,\omega_1 t + \frac{\sin\,\omega_1 t}{\sqrt{4Q^2 - 1}}\right),\label{eq:vstime}
\end{multline}
where $w_0/w_1=2Q/\sqrt{4Q^2 - 1}$.  The dashed line in Fig.~\ref{fig:finiteRise}(a) was generated using Eq.~(\ref{eq:vstime}) with the parameters \mbox{$f_0=\SI{42}{\mega\hertz}$} and $Q=0.6$.  It wasn't possible to find a combination of $f_0$ and $Q$ that captured both the rise time and the damped oscillations of the measured pulse.  The values used were chosen to match the initial rise of the measured pulse as closely as possible.  Figure~\ref{fig:finiteRise}(b) compares the measured and calculated transients responses when conductor losses and the finite rise time of the voltage pulse are included in the calculations.  The calculated response captures the detailed shape of the measured response over the entire range of measured times.  

Although we show the response only up to times of \SI{1}{\micro\second}, we have used INVLAP to numerically calculate $v_\mathrm{g}$ for times up to \SI{100}{\micro\second} following the voltage pulse and observed stable behavior at all times.  We also varied the values of $n_\mathrm{sum}$ and $n_\mathrm{dif}$ used when calling INVLAP.  Reducing both $n_\mathrm{sum}$ and $n_\mathrm{dif}$ by a factor of two has no perceptible effect on the calculated $v_\mathrm{g}$. 

\subsection{Open-circuit parasitic capacitance and oscilloscope input impedance}
The transient responses in Sections~\ref{sec:Ronly} and \ref{sec:finiteRise} assumed an infinite load impedance at the open end of the transmission line such that $Z_\mathrm{in}\approx Z_\mathrm{c}\coth\left(\gamma\ell\right)$.  However, fringing electric fields are known to result in an effective parasitic capacitance $C_\mathrm{p}$ terminating the line.\cite{Stuchley:1980, Stuchley:1982, Bobowski:2012a, Bobowski:2012b}  The parasitic capacitance was determined to be on the order of \SI{0.05}{\pico\farad} at the open end of a UT-141 coaxial cable.\cite{Stuchley:1982, Bobowski:2012b}  To include the effect of this parasitic capacitance in the $V_\mathrm{g}$ calculation, it is necessary to use the full expression for the transmission line input impedance given by Eq.~(\ref{eq:Zin}) with $Z_\mathrm{L}=\left(sC_\mathrm{p}\right)^{-1}$.  $\mathscr{L}^{-1}\{V_\mathrm{g}\}$ using $C_\mathrm{p}=\SI{0.05}{\pico\farad}$ was numerically evaluated and the extracted transient response was found to be nearly identical to response calculated in Sec.~\ref{sec:finiteRise} and shown in Fig.~\ref{fig:finiteRise}(b).

\begin{figure}
\centering{
\includegraphics[width=0.9\columnwidth]{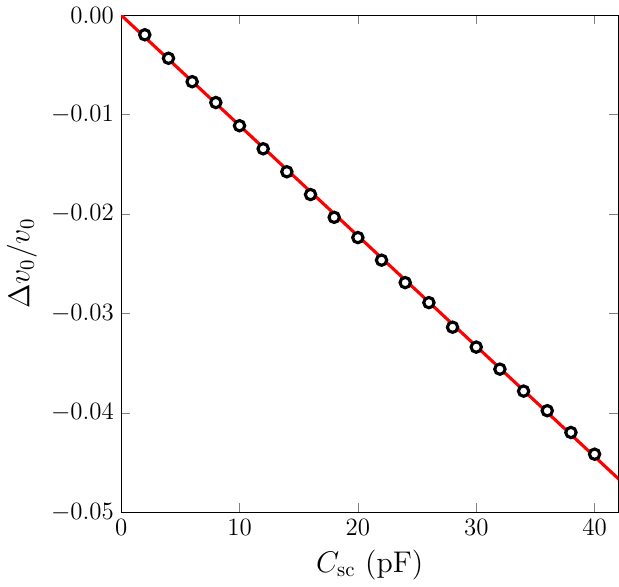}
}
\caption{\label{fig:v0Csc}Fractional error in the determination of $v_0$ when $C_\mathrm{sc}$ is neglected as a function of $C_\mathrm{sc}$.  The line is a linear fit to the data and has a slope of \SI{-0.0011}{\pico\farad\tothe{-1}}.
}
\end{figure}

\begin{figure*}[t]
\centering{
\begin{tabular}{cc}
(a)\includegraphics[height=0.87\columnwidth]{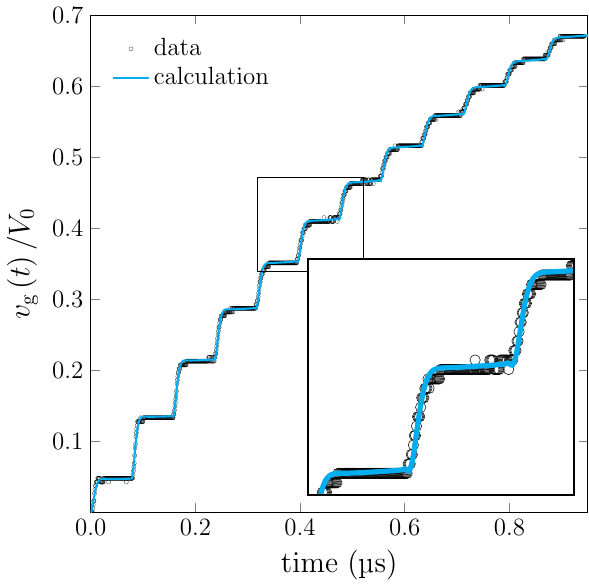} & \qquad (b)\includegraphics[height=0.85\columnwidth]{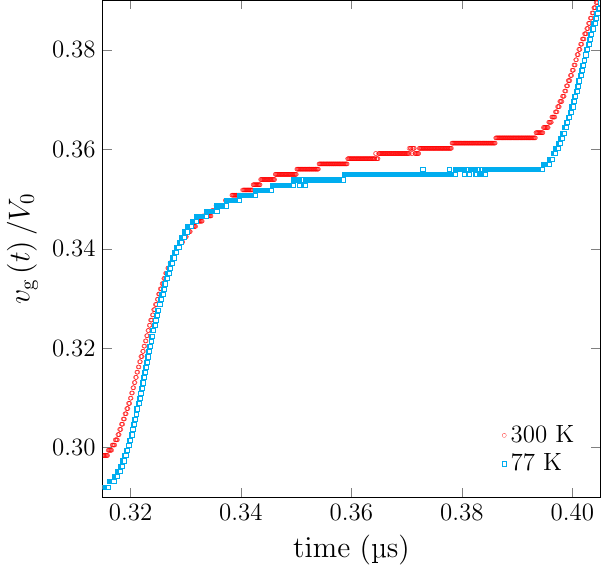}
\end{tabular}
}
\caption{\label{fig:LN2}(a) The measured and calculated transient response of the UT-141 coaxial cable at \SI{77}{\kelvin}.  The calculated response includes conductor losses, the finite rise time of the input voltage pulse, and the input impedance of the oscilloscope.  (b) A detailed comparison of the measured fifth step of $v_\mathrm{g}$ when the coaxial line is at room temperature and \SI{77}{\kelvin}.}
\end{figure*}

Finally, we used INVLAP to investigate the effect that the input impedance $Z_\mathrm{sc}$ of the oscilloscope has on the measured transient response.  The oscilloscope places a shunt resistance $R_\mathrm{sc}$ and a shunt capacitance $C_\mathrm{sc}$ in parallel with $Z_\mathrm{in}$ shown in Fig.~\ref{fig:transients}(b).  As a result, the effective impedance of the combination is determined from \mbox{$Z_\mathrm{eff}^{-1}=Z_\mathrm{in}^{-1}+R_\mathrm{sc}^{-1}+sC_\mathrm{sc}$} and $V_\mathrm{g}$ becomes
\begin{equation}
V_\mathrm{g}=V_\mathrm{s}\frac{Z_\mathrm{eff}}{R_\mathrm{g}+Z_\mathrm{eff}}.
\end{equation} 

The Tektronix TBS 1104 oscilloscope used to measure the transient response is specified to have $R_\mathrm{sc}=\SI{1}{\mega\ohm}$ and $C_\mathrm{sc}=\SI{20}{\pico\farad}$.  We first evaluated $\mathscr{L}^{-1}\{V_\mathrm{g}\}$ for many values of $R_\mathrm{sc}$ spanning \num{0.5} to \SI{2}{\mega\ohm} while keeping $C_\mathrm{sc}$ fixed and found that $v_\mathrm{g}$ has almost no dependence on the value of $R_\mathrm{sc}$.  On the other hand, scanning the value of $C_\mathrm{sc}$ while keeping $R_\mathrm{sc}$ fixed resulted in a noticeable change to the effective signal propagation speed.  Neglecting $C_\mathrm{sc}$ causes $v_0$ to be underestimated from the experimental data.  Figure~\ref{fig:v0Csc} shows the fractional error in the determination of $v_0$ that results from neglecting the effect of $C_\mathrm{sc}$.  $C_\mathrm{sc}=\SI{20}{\pico\farad}$ results in a $v_0$ value that is \SI{2.2}{\percent} below the true value.  The transient response $v_\mathrm{g}$ shown in Fig.~\ref{fig:finiteRise}(b) was generated using $v_0=0.680c$ and did not take into account the effects of $C_\mathrm{sc}$.  The response calculated when using $C_\mathrm{sc}=\SI{20}{\pico\farad}$ looks identical to that shown in Fig.~\ref{fig:finiteRise}(b) if $v_0=0.695c$ is used.  This result for the signal propagation speed is closer to the manufacturer specification of $v_0=0.7c$.\cite{MicroCoax}

\section{LOW-TEMPERATURE TRANSIENT RESPONSE}\label{sec:LN2}

In order to observe a transient response that more closely approximates that of an ideal transmission line, we cooled the UT-141 coaxial cable in liquid nitrogen and repeated the measurement shown schematically in Fig.~\ref{fig:transients}(a).  The \SI{8.07}{\meter} line was coiled, placed in a large polystyrene foam box, and then submerged in liquid nitrogen.  All but \SI{20}{\centi\meter} of the cable was completely below the surface of the nitrogen bath.  This short length was used to make a connection to $R_\mathrm{g}$ which remained at room temperature.  Cooling the transmission line reduces the resistivities of the cable's copper outer conductor and SPCW center conductor thereby reducing conductor losses.  Figure~\ref{fig:LN2}(a) shows the measured low-temperature transient response.  The cable was left submerged in nitrogen bath for approximately \SI{20}{\minute} before the data was acquired to allow thermal equilibrium to be established.  We also confirmed that the observed response was no longer evolving with time before recording the data.  Compared to the room temperature data shown in Fig.~\ref{fig:finiteRise}(b), the horizontal sections of the steps in the low temperature data have significantly shallower slopes. 
\begin{table}\caption{\label{tab:1}Parameters used to calculate the transient responses at room temperature and \SI{77}{\kelvin}.}
\begin{tabular}{llll}
\hline\hline\\ [-1.5ex]
\textbf{parameter}\qquad~~~~ & \textbf{\SI[detect-weight=true, detect-family=true]{300}{\kelvin}}\quad~~~ & \textbf{\SI[detect-weight=true, detect-family=true]{77}{\kelvin}}\quad~~~ & \textbf{property}\\[0.5ex]
\hline\\ [-1.5ex]
$v_0/c$ & \num{0.695} & \num{0.695} & characteristic\\
$Z_0$ (\si{\ohm}) & \num{49.05} & \num{48.20} & parameters\\[0.5ex]
\hline\\ [-1.5ex]
$a$ (\SI{e-4}{\ohm\,\second\tothe{1/2}/\meter}) & \num{1.252} & \num{0.250} & conductor losses\\[0.5ex]
\hline\\ [-1.5ex]
$f_0$ (\si{\mega\hertz}) & \num{42} & \num{42} & pulse rise\\
$Q$ & \num{0.6} & \num{0.6} & time\\[0.5ex]
\hline\\ [-1.5ex]
$R_\mathrm{sc}$ (\si{\mega\ohm}) & \num{1} & \num{1} & oscilloscope\\
$C_\mathrm{sc}$ (\si{\pico\farad}) & \num{20} & \num{20} & impedance\\[0.5ex]
\hline\hline
\end{tabular}
\end{table} 

The line in Fig.~\ref{fig:LN2}(a) was calculated using INVLAP and includes conductor losses, the finite rise time of the input voltage step, and the input impedance of the oscilloscope.  Table~\ref{tab:1} presents the parameters used to generate the final calculated transients responses at both room temperature and \SI{77}{\kelvin}.  At low temperatures, the $a$ coefficient characterizing conductor losses was reduced by a factor of five and we found that we had to lower $Z_0$ by \SI{1.7}{\percent} to account for a slight difference in the size of the voltage steps at the two temperatures.  For example, the second voltage step at \SI{85}{\nano\second} is \SI{87.3}{\milli\volt} at \SI{300}{\kelvin} and \SI{85.1}{\milli\volt} at \SI{77}{\kelvin}.  We note that we also calculated the low-temperature transient response while including a non-zero conductance due to dielectric losses, using Eqs.~(\ref{eq:Zc}) and (\ref{eq:gamma}) for $Z_\mathrm{c}$ and $\gamma$ rather than the low-loss approximations presented in Sec.~\ref{sec:Ronly}, and including a parasitic capacitance $C_\mathrm{p}$ terminating the line.  Even with reduced conductor losses, none of these changes, either on their own or in combination with one another, result in a significant change to the calculated response shown in Fig.~\ref{fig:LN2}(b).    

Figure~\ref{fig:LN2}(b) shows a detailed comparison of the fifth step of $v_\mathrm{g}$ at the two measurement temperatures.  The step rise times, controlled by the input voltage pulse, are very similar, but the slope of the regions between steps are clearly different due to the difference in conductor losses at the two temperatures.

\section{SUMMARY}\label{sec:summary}

We have used numerical inverse Laplace transforms to calculate the transient response of lossy transmission lines to a voltage step.  The calculations were compared to experimental measurements and found to be in good agreement when using realistic parameter values.  Conductor losses were found to have an important effect on the detailed shape of the transient response, whereas dielectric losses could be neglected.  Modeling the input voltage step as an underdamped oscillator allowed the finite rise time of the step to be incorporated into the calculation which was important for reproducing the finite slopes of the vertical steps in the measured transient response.  Finally, we found that it was important to include the input impedance of the oscilloscope in order to accurately determine the signal propagation speed.  Neglecting the \SI{20}{\pico\farad} input capacitance of the oscilloscope used in our measurements resulted in the propagation speed being underestimated by \SI{2.2}{\percent}.

%\section*{Acknowledgments}
%We thank the referees for their thorough review of the manuscript and insightful suggestions for improvements.


\begin{thebibliography}{99}
% The numeral (here 99) in curly braces is nominally the number of entries in
% the bibliography. It's supposed to affect the amount of space around the
% numerical labels, so only the number of digits should matter--and even that
% seems to make no discernible difference.

%1
\bibitem{Bobowski:2020}
J. S. Bobowski, ``Modeling and measuring the non-ideal characteristics of transmission lines,'' to appear in Am. J. Phys. \textbf{88}, (2020).

%2
\bibitem{Haus:1989}
H. A. Haus and J. R. Melcher, {\it Electromagnetic Fields and Energy}, (Prentice-Hall, New Jersey, 1989).

%3
\bibitem{Pozar:2012}
David M. Pozar, {\it Microwave Engineering}, 4th ed. (Wiley, New Jersey, 2012).

%4
\bibitem{Collier:2013}
Richard Collier, {\it Transmission Lines}, (Cambridge U.P., New York, 2013).

%5
\bibitem{Snoke:2015}
David W. Snoke, {\it Electronics: A Physical Approach}, (Pearson Education, Inc., Boston, 2015).


%6
\bibitem{Shenkman:2005}
A. L. Shenkman, {\it Transient Analysis of Electric Power Circuits Handbook}, (Springer, Dordrecht, NL, 2005).

%7
\bibitem{Griffith:1990}
J. R. Griffith and M. S. Nakhla, ``Time-domain analysis of lossy coupled transmission lines,'' IEEE Trans. Microw. Theory Tech., \textbf{38}, 1480--1487 (1990).

%8
\bibitem{Valsa:1998}
J. Valsa and L. Bran\v{c}ik, ``Approximate formulae for numerical inversion of Laplace transforms,'' Int. J. Numer. Modell. Electron. Networks Devices Fields, \textbf{11}, 153--166 (1998).

%9
\bibitem{Valsa:2011}
J. Valsa, ``Numerical Inversion of Laplace Transforms in Matlab'', \url{https://www.mathworks.com/matlabcentral/fileexchange/32824-numerical-inversion-of-laplace-transforms-in-matlab} [Accessed: Oct.~10, 2020], posted to the MathWorks File Exchange (2011).

%10
\bibitem{Stuchley:1980}
M. A. Stuchly and S. S. Stuchly, ``Coaxial line reflection methods for measuring dielectric properties of biological substances at radio and microwave frequencies - a review,'' IEEE Trans. Instrum. Meas. \textbf{29}, 176--183 (1980).

%11
\bibitem{Stuchley:1982}
M. A. Stuchly, T. W. Athey, G. M. Samaras and G. E. Taylor, ``Measurement of radio frequency permittivity of biological tissues with an open-ended coaxial line: Part II - experimental results,'' IEEE Trans. Microw. Theory Tech. \textbf{30}, 87--92 (1982).

%12
\bibitem{Bobowski:2012a}
J. S. Bobowski, T. Johnson, and C. Eskicioglu, ``Permittivity of waste-activated sludge by an open-ended coaxial line,'' Prog. Electromagn. Res. Lett. \textbf{29}, 139--149 (2012).

%13
\bibitem{Bobowski:2012b}
J. S. Bobowski and T. Johnson, ``Permittivity measurements of biological samples by an open-ended coaxial line,'' Prog. Electromagn. Res. B \textbf{40}, 159--183 (2012).

%14
\bibitem{MicroCoax}
{\it UT-141-HA-M17 Datasheet: Semi-rigid coaxial cable}, Micro-Coax, Pottstown, PA, USA.


\end{thebibliography}
\end{document}